\documentclass[pdflatex,sn-mathphys-num]{sn-jnl}

\usepackage{graphicx}%
\usepackage{multirow}%
\usepackage{amsmath,amssymb,amsfonts}%
\usepackage{amsthm}%
\usepackage{mathrsfs}%
\usepackage[title]{appendix}%
\usepackage{xcolor}%
\usepackage{textcomp}%
\usepackage{manyfoot}%
\usepackage{booktabs}%
\usepackage{algorithm}%
\usepackage{algorithmicx}%
\usepackage{algpseudocode}%
\usepackage{listings}%

\theoremstyle{thmstyleone}%
\theoremstyle{thmstyletwo}%

\theoremstyle{thmstylethree}%

\begin{document}

\title[Article Title]{Generation of high power spatially-structured laser pulses via forward Raman amplification in plasma}

\author[1,2]{\fnm{Zhiyu} \sur{Lei}}\email{zhiyulei@sjtu.edu.cn}

\author*[1,2]{\fnm{Suming} \sur{Weng}}\email{wengsuming@sjtu.edu.cn}

\author[1,2]{\fnm{Min} \sur{Chen}}\email{minchen@sjtu.edu.cn}

\author[1,2,3]{\fnm{Jie} \sur{Zhang}}\email{jzhang1@sjtu.edu.cn}

\author*[1,2,3]{\fnm{Zhengming} \sur{Sheng}}\email{zmsheng@sjtu.edu.cn}

\affil[1]{\orgdiv{State Key Laboratory of Dark Matter Physics, School of Physics and Astronomy}, \orgname{Shanghai Jiao Tong University}, \orgaddress{\city{Shanghai}, \postcode{200240}, \country{China}}}

\affil[2]{\orgdiv{Key Laboratory for Laser Plasmas and Collaborative Innovation Centre of IFSA}, \orgname{Shanghai Jiao Tong University}, \orgaddress{\city{Shanghai}, \postcode{200240}, \country{China}}}

\affil[3]{\orgdiv{Tsung-Dao Lee Institute}, \orgname{Shanghai Jiao Tong University}, \orgaddress{\city{Shanghai}, \postcode{201210}, \country{China}}}

\date{\today}

\abstract{Spatially-structured light with tunable intensity, wavelength, and spatiotemporal profiles has demonstrated significant potentials for fundamental and applied science, including the ultrafast and high-field physics. Nevertheless, the generation or amplification of such light towards extremely high power remains challenging due to the limitations of conventional gain media. Building upon our recently proposed forward Raman amplification (FRA) mechanism [Lei et al., Phys. Rev. Lett. 134, 255001 (2025)], here we develop a universal plasma-based amplification scheme that is capable of generating high-power structured laser beams, including vortex, Bessel, and Airy beams. Through theoretical modeling and multi-dimensional particle-in-cell simulations, we demonstrate that a near-infrared structured seed laser with an initial intensity of 10¹² W/cm² can achieve $10^4–10^5$-fold intensity amplification via FRA, and subsequently be self-compressed to sub-cycle duration with petawatt-level peak power. Benefiting from its exceptionally high amplification growth rate, the FRA process requires only femtosecond-scale interaction time and submillimeter propagation distance in plasma, effectively suppressing concomitant plasma instabilities. The high output intensity ($10^{17}\,\mathrm{W/cm^2}$), compactness ($<500\,\mathrm{\mu m}$), high temporal contrast, universal applicability to diverse structured beams, and relatively easy implementation with the co-propagating configuration combine to make the FRA a disruptive approach to the generation of petawatt-class spatially-structured light, enabling unprecedented applications in high-field physics and ultrafast science.}


\maketitle

\newpage
\section{Introduction}

Since the laser was invented in 1960~\cite{maiman1960stimulated}, it has become an essential tool for various applications in industry, medicine, and fundamental science owing to its high coherence and high brightness. Fundamentally, a laser beam can have certain spatial structure~\cite{rubinsztein2016roadmap,forbes2021structured,piccardo2025trends}, which provide a degree of freedom on light field modulation through precise control of amplitude, phase, or polarization distribution beyond conventional Gaussian beams. The complex and unique wavefront structures owned by such spatially-structured light bring some intriguing characteristics, such as diffraction-free propagation for a Bessel-Gaussian (BG) beam~\cite{durnin1987diffraction}, self-healing and self-acceleration pattern for an Airy beam~\cite{broky2008self,siviloglou2008ballistic}, orbital angular momentum (OAM) carried by a vortex beam or Laguerre-Gaussian (LG) beam ~\cite{allen1992orbital,shen2019optical}, and spin angular momentum (SAM) coupling in a vector beam~\cite{holleczek2011classical,rosales2018review}.
Over the past decades, spatially-structured lights have drawn tremendous attention and applied in numerous fields spanning from optical communication~\cite{wang2012terabit,bozinovic2013terabit}, atomic manipulation~\cite{padgett2011tweezers}, to high resolution imaging~\cite{torner2005digital}. In the high-intensity region, spatially-structured light like LG beam has been applied for laser-driven particle acceleration~\cite{vieira2014nonlinear,vieira2018optical,wang2025enhanced}, strong transient magnetic field generation~\cite{shi2018magnetic,longman2021kilo}, and high-order harmonic generation (HHG)~\cite{chen2021intense,zhang2015generation}. Recently, numerical simulations have shown that the intense BG beam and Airy beam are promising to produce high-intensity HHG and isolated attosecond pulse~\cite{pang2024self,chen2024isolated}. While the high power spatially-structured lasers have immense application potential, experimental realization and practical implementation of these applications remain fundamentally constrained by persistent barriers in generating high-quality intense spatially-structured light beams.

Currently, the generation of spatially-structured laser beams is mainly based on solid-state optical elements such as phase plates~\cite{beijersbergen1994helical}, q-plates~\cite{marrucci2006optical}, conical mirrors~\cite{kuntz2009spatial}, or customized spatial light modulators. The resultant laser beams then can be amplified via chirped pulse amplification (CPA)~\cite{strickland1985compression} or the optical parametric chirped-pulse amplification (OPCPA)~\cite{ross1997prospects}. However, further enhancement on spatially-structured laser intensity is hindered by inherent low damage threshold ($\sim 1\,\mathrm{J/cm^2}$@$10\,\mathrm{ps}$) of the involved solid-state optical elements mentioned above~\cite{stuart1995laser}. Meanwhile, the output wavelengths of spatially-structured laser pulses remain restricted to around $0.8\,\mathrm{\mu m}$ (Ti:Sapphire) and $1.06\,\mathrm{\mu m}$ (Nd:YAG) due to currently available gain media. In recent years, plasma-based schemes for the manipulation and amplification of high-power lasers have attracted increasing attention~\cite{hur2023laser,nie2018relativistic,zhu2020efficient,riconda2023plasma}. Since the maximum energy density the plasma can sustain is several orders of magnitude higher ($\sim 10^{17}\,\mathrm{W/cm^2}$) than that of the solid crystal ($\sim 10^{13}\,\mathrm{W/cm^2}$)~\cite{stuart1996optical}, plasma-based amplification schemes have the potential to generate even more powerful laser pulses than those based purely on solid-state optical elements. So far, three-wave coupling has been considered as the most effective light amplification mechanism in plasma, including processes such as stimulated Raman scattering (SRS)~\cite{malkin1999fast,trines2011simulations,turnbull2018raman,cheng2005reaching,ren2007new} and the stimulated Brillouin scattering (SBS)~\cite{andreev2006short,riconda2013spectral,weber2013amplification,lancia2016signatures,marques2019joule}. As long as the phase-matching condition is satisfied, the excited plasma wave (Langmuir electron wave for SRS or ion acoustic wave for SBS) can continuously scatter the energy from the long, intense pump pulse to the short, weak seed pulse. Recent numerical studies have shown that both backward SRS~\cite{vieira2016amplification} and backward SBS schemes~\cite{wu2024efficient} can be used to amplify vortex beams in plasma. Currently, both SRS and SBS amplification schemes are primarily triggered by backward scattering, where the pump and seed pulses are counter-propagating in plasma and undergo a head-on collision. Typically, these backward-scattering configurations require conditions such as a relatively long plasma length (mm scale), long interaction time (ps scale), low plasma density, and a specific angle between the two pulses. These requirements result in a low amplification growth rate, low energy conversion efficiency, the emergence of significant kinetic instabilities, and the complicated experimental design for spatial-temporal alignment. Moreover, only seed pulses with wavelength close to or matching the pump pulse’s wavelength can be amplified effectively. These limitations pose major obstacles to the backward-scattering amplification schemes. While theoretical models predict the production of PW-class vortex beams with 1 mm spot sizes, no plasma-based Raman amplification beyond $0.1\,\mathrm{TW}$ has been experimentally demonstrated to date~\cite{trines2020new,shi2024advances}.

Recently, we proposed a new class of plasma amplification mechanisms via forward Raman scattering, i.e., forward Raman amplification (FRA)~\cite{cshj-jgz7}, where the seed and pump pulses co-propagate in a moderate density plasma. In this paper, we further propose to apply the FRA scheme to scenarios involving spatially-structured laser beams, including LG, BG, and Airy beams. Here, the seed pulse is a spatially-structured laser beam, while the pump laser is a conventional Gaussian beam. An analytical theory has been developed to illustrate the capability of the FRA scheme for arbitrary transverse laser field profiles of the seed pulse. For the LG beam amplification, the phase-matching condition for the FRA including the generation of new OAM mode is theoretically derived and numerically verified. Three-dimensional particle-in-cell (PIC) simulations further demonstrate that the FRA scheme can effectively achieve intensity amplification factors of $10^4 \sim 10^5$ for the aforementioned spatially-structured lasers. The FRA processes of these spatially-structured beams are consistent with  that of conventional Gaussian pulses, and thus they can be well predicted by the universal theoretical models.

\section{Scheme and Theoretical Model}\label{sec_m}
\begin{figure}[t]
\centering
\includegraphics[width=0.9\textwidth]{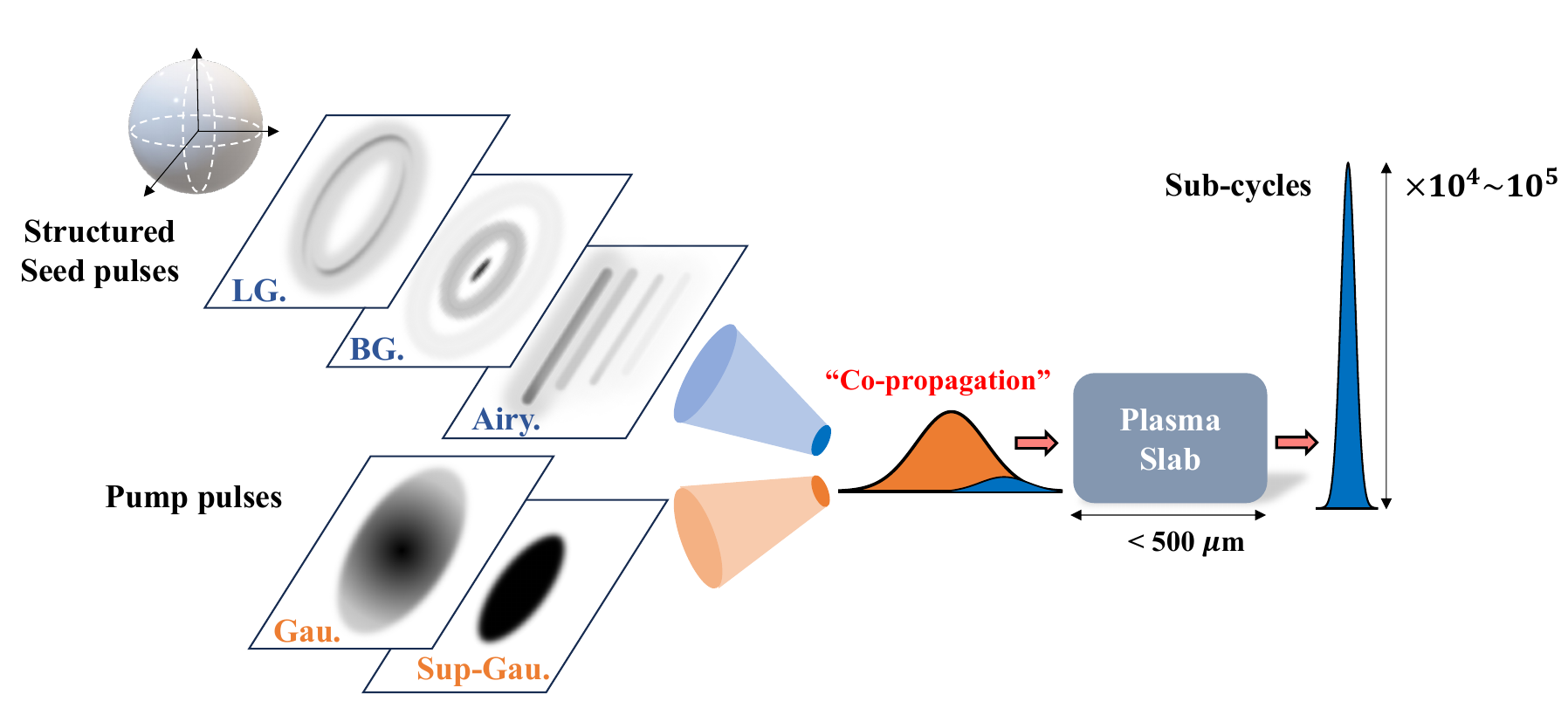}
\caption{\textbf{Schematic diagram for the amplification of spatially-structured laser beams via FRA.} The seed pulse (blue) can be arbitrary spatially-structured light, including LG beam, BG beam, and Airy beam. The pump pulse (red) can be either a general Gaussian pulse or a super-Gaussian pulse. The two pulses co-propagate in the forward direction and inject into the uniform plasma slab (few hundreds of micrometers length only), with the seed pulse located at the front of the pump pulse at the beginning. The output seed pulse can achieve $10^4\sim 10^5$ times of amplification and its duration is further self-compressed to sub-cycles. }
\label{model}
\end{figure}

The FRA process of a laser involves three stages. Initially, an intense pump laser (frequency $\omega_0$ , wavenumber $k_0$) with a longer duration and shorter wavelength than the seed pulse to be amplified is incident together with the seed ($\omega_1$, $k_1$) along the same direction into a homogeneous plasma with a moderately high density. The laser-plasma interaction excites an electron plasma wave characterized by frequency $\omega_2\approx\omega_{\mathrm{pe}}$ and wavenumber $k_2$, respectively, where $\omega_{\mathrm{pe}}$ is the electron plasma frequency. The following phase-matching conditions are fulfilled for forward Raman scattering
\begin{equation}
\omega_0=\omega_1+\omega_2,\,\,\,k_0=k_1+k_2.
\end{equation}
Next, the pump laser (with a higher group velocity in plasma) catches up to and overtakes the seed pulse. During their co-propagation, significant energy transfer from the pump to the seed takes place via the forward Raman scattering. In this process, the electron plasma wave is efficiently excited by the beat wave of two pulses, which continuously mediates the energy transfer from the pump to the seed. Meanwhile, due to early pump depletion, the pump is unable to further amplify the leading edge of the seed pulse once it overtakes the seed. As a result, only a localized region of the seed pulse undergoes effective amplification, leading to a “compression”  of the seed pulse duration. If the seed pulse is an initially narrow one, however, its duration will be stretched. At the end, when the amplified seed pulse is intense enough, it will undergo self-compression to nearly single-cycle duration via the self-phase modulation (SPM) process. As shown in Fig.~\ref{model}, in the FRA scheme, the transverse profile of the pump pulse can be a Gaussian or super-Gaussian distribution, and the seed pulse can be either a regular Gaussian beam or various spatially-structured beams such as vortex, Airy, Bessel, Hermite or vector beams. The FRA scheme can achieve an intensity amplification factor of $10^4\sim 10^5$ for spatially-structured lasers within small spatial and temporal scales (hundreds of micrometers in space and hundreds of femtosecond in time). These features may enable the FRA scheme to be experimentally demonstrated under more accessible laboratory conditions.

To achieve the above process efficiently, one of the most important conditions is the use of moderately high density plasma. On one hand, it ensures sufficient group velocity difference between the pump and seed pulses when they co-propagate in the plasma (i.e. $v_{g0,g1}=c\sqrt{1-\omega_{\mathrm{pe}}^2/\omega_{0,1}^2}$), so that the seed pulse can continuously interact with the non-depleted portion of the pump for sustained amplification. On the other hand, both the amplification growth rate and  energy transfer efficiency from the pump increase with the plasma density. Moreover, the high-density plasma allows an efficient compression of the amplified seed pulse in the later stage. The above analysis indicates that the seed pulse to be amplified should have a longer wavelength to satisfy the three-wave resonant coupling condition in plasma, i.e., $\omega_{1}\approx\omega_{0}-\omega_{\mathrm{pe}}$, where the electron plasma frequency $\omega_{\mathrm{pe}}$ is directly related to the plasma electron density $n_e$ as $\omega_{\mathrm{pe}}^2 = 4\pi n_e e^2/m_e$. Therefore, the seed laser wavelength is determined by the pump wavelength and plasma density, given by $\lambda_1=\lambda_0/(1-\sqrt{n_e/n_{c,0}})$. Theoretically, the seed pulse with any given wavelength between $\left[ \lambda_{0},\,2\lambda_{0}   \right]$ can be amplified by this FRA scheme with a matched plasma density. In this paper we specifically focus on the long-wavelength regime to produce high-power near-infrared pulses.

The theoretical model of FRA with spatially-structured light beams is given below. Considering a configuration where the pump and seed lasers co-propagate in a homogeneous plasma slab along the positive $x$-direction. The spatiotemporal evolution of the pump, seed, and electron plasma wave is then governed by the following three-wave coupling equations:
\begin{align}
(\frac{\partial^2}{\partial t^2}-c^2\nabla^2+\omega_{\mathrm{pe}}^2)\bar{A_0}&=-\omega_{\mathrm{pe}}^2\frac{\delta {n}}{n_0}\bar{A_1}, \label{3wave0}\\
(\frac{\partial^2}{\partial t^2}-c^2\nabla^2+\omega_{\mathrm{pe}}^2)\bar{A_1}&=-\omega_{\mathrm{pe}}^2\frac{\delta {n}}{n_0}\bar{A_0}, \label{3wave1}\\
(\frac{\partial^2}{\partial t^2}-3v_{\mathrm{th}}^2\nabla^2+\omega_{\mathrm{pe}}^2)\frac{\delta {n}}{n_0}&=\frac{e^2}{m_e^2c^2}\nabla^2(\bar{A_0}\cdot\bar{A_1}). \label{3wave2}
\end{align}
Here $\bar{A}_{0,1}$ are the vector potentials of the incident and scattering lights, respectively, and $\delta n$ is the plasma density perturbations driven by the electron plasma wave. These three components satisfy the three-wave coupling conditions for Raman scattering. It is noteworthy that in the FRA scheme, the pump laser corresponds to the incident light in the forward Raman scattering process, while the seed laser represents the scattered light. Assuming the incident and scattered lights are propagating along x direction with linear polarization, we introduce $\bar{A}_{0,1}=\bar{A}_{0,1}^{'}\mathrm{exp}(ik_{0,1}x-i\omega_{0,1}t)+c.c.$, and write the plasma electron density perturbation as $\delta n/n_0=\delta n'\mathrm{exp}(ik_2-i\omega_2 t)+c.c.$. Using the slowly varying envelope approximation and considering the cold plasma case, Eqs.~\eqref{3wave0}$\sim$\eqref{3wave2} then become
\begin{align}
D_0\bar{A}_{0}^{'}=\omega_{\mathrm{pe}}^2\delta n'\bar{A}_{1}^{'}&,\,\,\,
D_1\bar{A}_{1}^{'}=\omega_{\mathrm{pe}}^2\delta n'^{*}\bar{A}_{0}^{'},\\
D_2\delta n'&=\frac{e^2k_2^2}{m_e^2c^2}\bar{A}_{0}^{'}\bar{A}_{1}^{'*},
\end{align}
where operators are defined as $D_{j}=2i\omega_{j}\partial_t+c^2(2ik_{j}\partial_x+\nabla_{\bot}^2)$ ($j=0,1$), $D_2=2i\omega_{\mathrm{pe}}\partial_t$. To simplify above equations, we  adopt the decomposition that the vector potential envelopes of pump and seed pulses $\bar{A}_{j}^{'}$ can be expressed as
\begin{equation}
\bar{A}_{j}^{'}=\bar{A}_{\parallel,j}(x,t)T_{j}(x,\bar{r}_{\bot}),
\end{equation}
where $\bar{A}_{\parallel}$ corresponds to the longitudinal envelope profile, $T$ corresponds the transverse envelope profile, and $\bar{r}_{\bot}$ represents the transverse coordinate. Similarly, the electron density perturbation can be expressed as $\delta n'=\delta n_{\parallel}(x,t)T_{2}(x,\bar{r}_{\bot})$. Next, we assume that the transverse envelopes obey to the paraxial approximation, which leads to $c^2(2ik_{j}\partial/\partial x+\nabla_{\bot}^2)T_{j}\approx 0$, so that
\begin{align}
D_{j}\bar{A}_{j}^{'}\approx2iT_{j}(\omega_{j}\partial_t+c^2k_{j}\partial_x)\bar{A}_{\parallel,j}.
\end{align}
Here we assume that the spots of the pump and seed pulses are unchanged during the propagation (i.e.$\partial_{x,t} T=0$), and their longitudinal profiles are independent of the transverse dimensions (i.e.$\nabla^2_{\bot}\bar{A}_{\parallel} =0$). Introducing the normalization $A_{j}^{'}=(m_ec^2/2e) a_j$, $\delta n'=ick_2/2\omega_{\mathrm{pe}}\sqrt{\omega_0/\omega_{\mathrm{pe}}}a_2$, and the three-wave coupling equations are then simplified as
\begin{align}
(\partial_t+v_{0}\partial_x)a_0&=\frac{ck_2}{4}\sqrt{\frac{\omega_{\mathrm{pe}}}{\omega_0}}a_1a_2,\\
(\partial_t+ v_{1}\partial_x)a_1&=-\frac{ck_2}{4}\sqrt{\frac{\omega_{\mathrm{pe}}\omega_0}{\omega_1^2}}a_0a_2^*\label{3w},\\
\partial_t a_2&=-\frac{ck_2}{4}\sqrt{\frac{\omega_{\mathrm{pe}}}{\omega_0}}a_0a_1^*,\label{3w_a2}
\end{align}
where $v_j=d\omega_j/dk_j=c^2/(\omega_j/k_j)$ is the group velocity of the laser pulse, determined by the electromagnetic wave dispersion relation in plasma $\omega_j^2=c^2k_j^2+\omega_{\mathrm{pe}}^2$.

The above equations are exactly the same as the coupling equations of forward Raman scattering in one-dimensional case, as shown in our previous work. It suggests that the FRA scheme is independent of $T(x,\bar{r}_{\bot})$. In other word, the amplification of spatially structured pulses with various transverse profiles (such as LG, BG, and Airy beams) exhibit identical laws in both linear and nonlinear stages as those governing conventional Gaussian beam amplification. Consequently, the analytical models that we derived in our previous work can be directly applied for the FRA of spatially structured lights. In the linear stage, the pump depletion is negligible, and $a_0 = a_{00}$ is a constant. The three-wave coupling equation then can be solved analytically~\cite{cshj-jgz7}
\begin{equation}
a_1=a_{10}I_0(2g\sqrt{\zeta\tau}),
\label{l_model}
\end{equation}
where $\tau=x/v_1,\,\,\,\zeta=t-\tau$ are the space and time variables in the co-moving coordinate system, and $a_{10}$ is the seed initial amplitude. The linear growth rate of FRA is defined by $g=a_{00}ck_2\sqrt{\omega_{\mathrm{pe}}/(\omega_0-\omega_{\mathrm{pe}})}/4$, with $k_2\approx(\sqrt{\omega_0^2-\omega_{\mathrm{pe}}^2}-\sqrt{\omega_0^2-2\omega_0\omega_{\mathrm{pe}}})/c$ determined by the linear dispersion relations of three waves for the Raman forward scattering. For the nonlinear stage,  the scaling relation governing the seed pulse intensity has been established and can be expressed as~\cite{cshj-jgz7}
\begin{equation}
a_1\approx a_{00}^2 a_{10}\delta\tau\omega_0/(\omega_0-\omega_{\mathrm{pe}}).\label{nl_model}
\end{equation}
Here the space and time coordinates $(\tau,\zeta)$ are multiplied by a normalized factor $ck_2\sqrt{\omega_{\mathrm{pe}}/\omega_0}$, and $\delta$ is the seed pulse duration. It should be mentioned that the above results are obtained under the assumption of paraxial approximation, which requires the propagation distance to be much smaller than the Rayleigh length. Compared with the amplification schemes based on backward scattering, the FRA can achieve significant amplification within a shorter propagation distance in the plasma. This implies that the FRA inherently satisfies the paraxial approximation assumption, and hence are appropriate for the effective amplification of spatially-structured laser beams.

\section{Simulation Results}

\subsection{Laguerre-Gaussian Beams}

Let us first consider the amplification of a LG vortex pulse via FRA. For a linearly polarized LG pulse propagating along $+x$ direction, the electric field is given by~\cite{allen1992orbital}
\begin{align}
a&_{\mathrm{LG}}=a_0C_p^l\frac{\sigma_0}{\sigma(x)}\left(\frac{\sqrt{2}r}{\sigma(x)}\right)^{|l|}L_{p}^{|l|}(-\frac{2r^2}{\sigma^2(x)})\mathrm{exp}(il\theta)\nonumber\\
&\mathrm{exp}\left(-\frac{r^2}{\sigma^2(x)}+i\frac{kr^2}{x^2+x_R^2}x-i\zeta(x)+\phi_0(x,t)\right)
\end{align}
where $p$ and $l$ are the respective radial mode index and the topological charge, $a_0$ the initial normalized laser amplitude, $C_p^l$ is the normalizing constant, $\sigma_0$ is the spot size at focus, $\sigma(x)=\sigma_0\sqrt{1+(x/x_R)^2}$ is the waist of the beam as a function of the propagation distance, $x_R=\pi\sigma_0^2/\lambda$ is the Rayleigh length, $\lambda$ is the central wavelength, $r=\sqrt{y^2+z^2}$ is the radial distance to axis, $L_p^{|l|}$ is the generalized Laguerre polynomial, $\theta=\arctan(z/y)\in \left[0,\,2\pi \right]$ is the azimuthal angle, $\zeta=(l+2p+1)\arctan(x/x_R)$ is the Gouy phase, and $\phi_0=kx-\omega t$ is the propagation phase of the LG pulse. Since the FRA inherently satisfies the paraxial approximation, we can then assume that $x\ll x_R$. In this section, we only consider the LG modes with $p=0$, so that the transverse profile of the LG beam can be then simplified as $T\approx (\sqrt{2}r/\sigma_0)^l\mathrm{exp}(-r^2/\sigma_0^2+il\theta)$.

One of the most remarkable features of the LG pulse is that an extra phase term $\mathrm{exp}(il\theta)$ is introduced in the electric fields, which corresponds to the orbital angular momentum (OAM) that the LG beam carries and is quantified by the topological charge $l$. Therefore, the rules governing the angular momenta of three coupling waves in forward Raman scattering must be considered. Given that all the phase factors in the three-wave coupling equations shall cancel out, the topological charges of the three waves are conserved. The complete phase matching conditions for the LG beams in forward Raman amplification then become
\begin{equation}
\omega_0=\omega_1+\omega_2,\,\,\,k_0=k_1+k_2,\,\,\,l_{0}=l_{1}+l_{2},
\label{phase_match_LG}
\end{equation}
which represent the conservations of wave energy, linear momentum, and orbital angular momentum, respectively. Now considering that both the pump and seed pulses have OAM components in transverse directions $y$ and $z$, the initial electric field can then be described as $\bar{a}_{0j}=a_{jy}\mathrm{exp}(il_{jy}\theta)\hat{y}+ia_{jz}\mathrm{exp}(il_{jz}\theta)\hat{z}$, where $j=0,\,1$ corresponds to the pump and seed pulses, respectively. From the coupling equations governing the evolution of the electron plasma wave (Eq.~\eqref{3w_a2}) and the seed pulse in the nonlinear stage (Eq.~\eqref{nl_model}), one can obtain
\begin{align}
\delta n_e&\sim \bar{a}_{00}\bar{a}_{01}^{*}\nonumber\\
&\propto a_{0y}a_{1y}^{*}e^{ i(l_{0y}-l_{1y})\theta}+ a_{0z}a_{1z}^{*}e^{ i(l_{0z}-l_{1z})\theta}\label{ne_OAM}\\
\bar{a}_1&\sim \bar{a}_{10}\bar{a}_{00}^{*}\bar{a}_{00}\nonumber\\
&\propto (a_{1y}a^2_{0y}e^{il_{1y}\theta}+a_{1z}a_{0z}^{*}a_{0y}e^{i(l_{1z}-l_{0z}+l_{0y})\theta})\hat{y}\nonumber\\
&+i(a_{1z}a^2_{0z}e^{il_{1z}\theta}+a_{1y}a_{0y}^{*}a_{0z}e^{i(l_{1y}-l_{0y}+l_{0z})\theta})\hat{z}.\label{a1_OAM}
\end{align}
One can draw the following conclusions from the above analysis. \textbf{(1)} The electron plasma wave driven by the beat wave of pump and seed pulses during the FRA may acquire OAM if the pump or seed pulse carries OAM. For example, when a linearly polarized Gaussian pump beam  ($l_{0y}=l_{0z}=0,\,a_{0z}=0$) is applied to amplify a linearly polarized seed LG beam ($l_{1y}=l,\,l_{1z}=0,\,a_{1z}=0$), the electron plasma wave will acquire the OAM with the topological charge $l_2=l_{0y}-l_{1y}=-l$ according to Eq.~\eqref{ne_OAM} , which is also consistent with the results given by phase matching condition (Eq.~\eqref{phase_match_LG}). \textbf{(2)} The amplified seed pulse can generate new OAM modes. For example, when the pump is a circularly polarized Gaussian beam ($l_{0y}=l_{0z}=0$), and the initial seed pulse is a linearly polarized LG beam ($l_{1y}=l,\,a_{1z}=0$), Eq.~\eqref{a1_OAM} suggests that the amplified seed pulse will have two OAM modes after the FRA. One is the original mode $l_{1y}=l$ along $y-$direction, and the other one is the newly generated mode $l_{1y}-l_{0y}+l_{0z}=l$ along $z-$direction. The new mode satisfies the OAM matching condition as well. Initially, the electron plasma wave obtains the OAM mode $l_2=l_{0y}-l_{1y}$ since both the pump and seed pulses have electric field components along $y-$direction, and the three-wave coupling will be triggered. Meanwhile, since the plasma wave is a longitudinal wave, its OAM shall be conserved in all transverse directions including the $z-$direction. Therefore, the same plasma wave will then couple with the electric field component of the pump along the $z-$direction, and the conservation of OAM ensures that the seed will obtain the additional OAM mode in the $z-$direction with $l_{1z}=l_{0z}-l_2=l_{0z}-l_{0y}+l_{1y}$.

We have performed a series of three-dimensional (3D) PIC simulations using the EPOCH code~\cite{arber2015contemporary} with a moving window technique to demonstrate the amplification of LG beam via the FRA scheme. The input seed pulse is a $y$-linearly polarized LG beam carrying OAM ($l_{1y}=1$), with wavelength $\lambda_1=1.8\,\mathrm{\mu m}$, peak intensity $I_1=1\times 10^{12}\,\mathrm{W/cm^2}$, duration $\tau_1=90\,\mathrm{fs}$, and spot size $\sigma_1=40\,\mathrm{\mu m}$. The pump pulse is a $y$-linearly polarized super-Gaussian beam without OAM ($l_{0y}=0$), with wavelength $\lambda_0=1.0\,\mathrm{\mu m}$, peak intensity $I_0=2\times 10^{16}\,\mathrm{W/cm^2}$, duration $\tau_0=180\,\mathrm{fs}$, and spot size $\sigma_0=40\,\mathrm{\mu m}$. Both the pump and the seed pulses have a $\sin^2$ temporal profile, and co-propagate along the $x-$direction. The plasma is uniformly distributed along the $x-$direction with a density of $2.2\times 10^{20}\,\mathrm{cm^{-3}}$ to satisfy the three-wave coupling conditions for forward Raman scattering. The advantage of employing a super-Gaussian pump is its potential to achieve a higher amplification efficiency within a limited spot size. While some previous studies suggest that super-Gaussian pumps may lead to stronger filamentation of the seed pulse~\cite{li2018focusing}, it has been demonstrated that light beams carrying OAM can propagate over longer distances in underdense plasma without significant filamentation~\cite{ju2016controlling}.
\begin{figure}[t]
\centering
\includegraphics[width=1.0\textwidth]{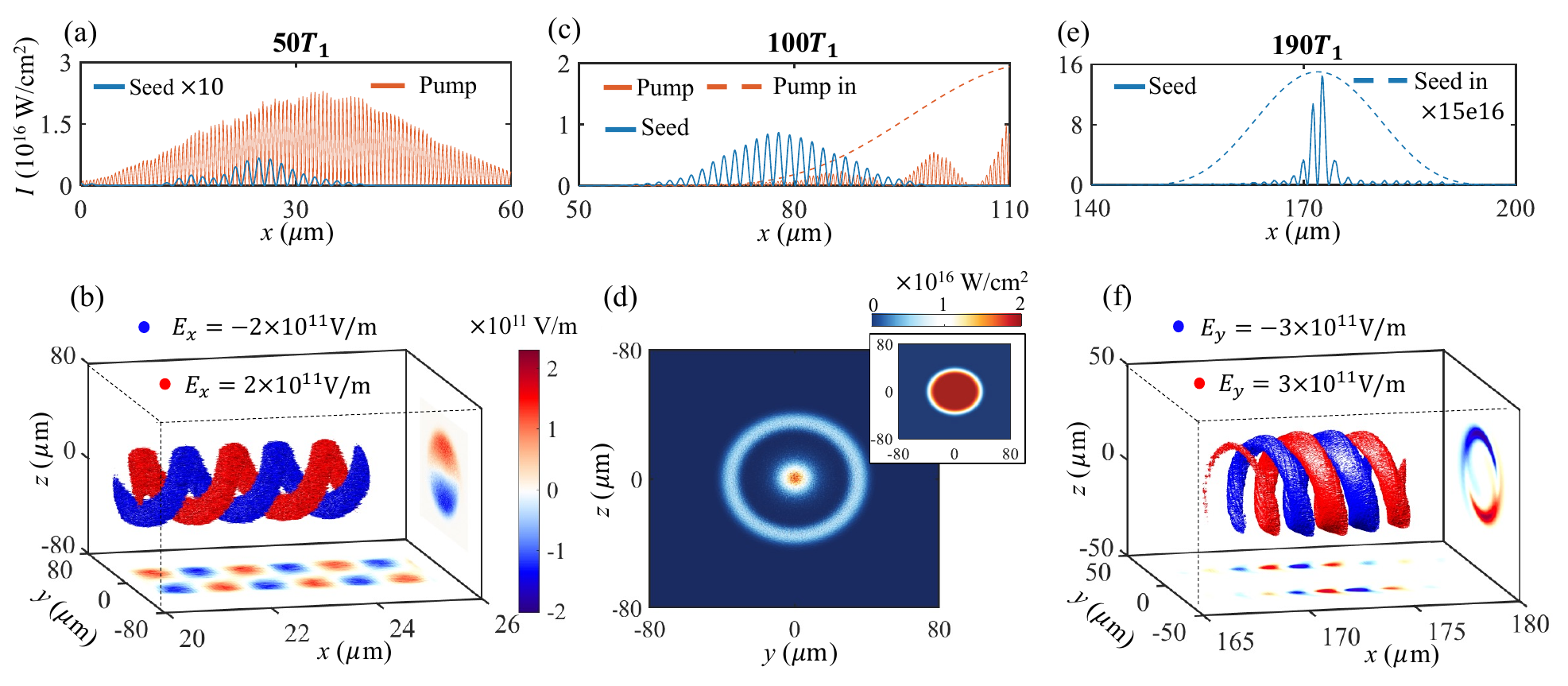}
\caption{\textbf{3D PIC Simulation results showing the amplification of a LG seed beam by a super-Gaussian pump pulse.} (a)(c)(e) The selected 1D spatial distributions of the seed (blue) and pump pulse (red) intensity at the early stage ($50\,T_1$), middle stage ($100\,T_1$), and final stage ($190\,T_1$), where $T_1$ is the seed pulse cycle. Seed intensity $\times 10$ magnified in (a) to make it visible; The red dashed line in (c) is the initially longitudinal envelope of the pump; The blue dashed line in (e) is the initially longitudinal envelope of the seed magnified by $1.5\times 10^{17}$. (b) The 3D isosurfaces of the electrostatic field $E_x$ at the early stage. (d) The 2D slice of pump transverse intensity distributions at the middle stage, and the insert one is the initial pump without depletion. (f) The 3D isosurfaces of the LG seed pulse electric field $E_y$ at the final stage.}
\label{LG}
\end{figure}

\begin{figure}[b]
\centering
\includegraphics[width=0.65\textwidth]{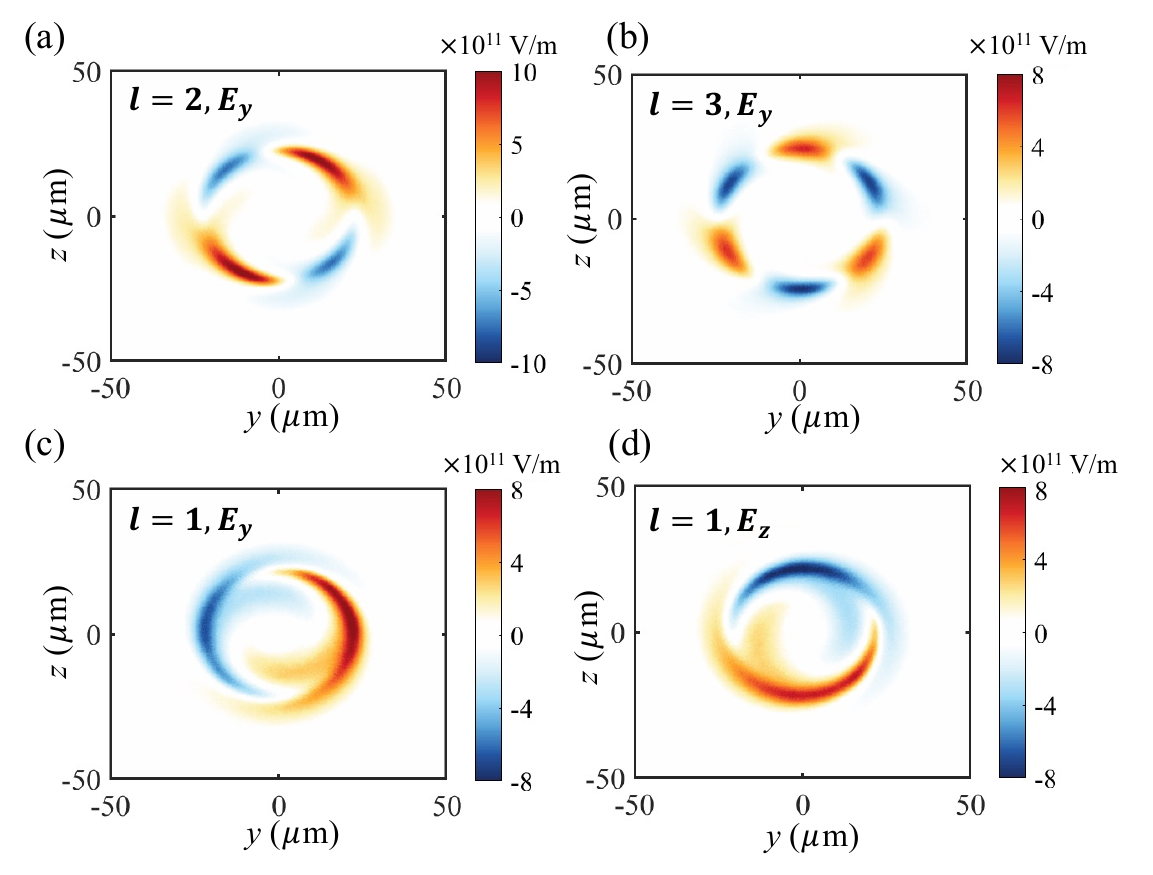}
\caption{\textbf{The amplification on LG beams with different OAM}, where the transverse electric field distributions of the seed LG beam are given by 3D PIC simulations. (a)(b) A LG beam linearly-polarized along the $y-$direction with $l_{1y}=2$ and $l_{1y}=3$, amplified by a linearly polarized pump. (c)(d) A LG beam linearly-polarized along the $y-$direction with $l_{1y}=1$, amplified by a circularly polarized pump, including the existing component $E_y$ and the new component $E_z$.}
\label{LG_l}
\end{figure}

With the setup above, the 3D PIC simulation results are shown in the Fig.~\ref{LG}. In the early stage ($\sim 50T_1$), as can been seen from the electric field distribution of the pulses (Fig.~\ref{LG}(a)), the pump pulse is barely depleted, while the seed pulse is preliminarily amplified to around $10^{14}\,\mathrm{W/cm^2}$, corresponding to the linear stage of FRA. Meanwhile, within the overlapping region of the two pulses, electrons in the background plasma are strongly resonant due to the three-wave coupling. The resultant  electron plasma wave (i.e. electron density perturbation $\delta n_e$), which can be represented by the electrostatic field $E_x$, shows a feature of helical structure with only two-petal transverse profile (Fig.~\ref{LG}(b)). Such phenomenon indicates that the electron plasma wave has acquired the OAM with a topological charge of $l_2=-1$, which is consistent with our theoretical prediction by Eq.~\eqref{ne_OAM}.
As the two pulses evolve into the middle stage ($\sim 100T_1$), the pump gradually overtakes the seed due to its higher group velocity in the plasma. During this stage, the seed pulse intensity further increases beyond $10^{16}\,\mathrm{W/cm^2}$, accompanied by significant pump depletion. In the Fig.~\ref{LG}(c), the longitudinal field distribution of the pump pulse transforms into a wave train, indicating the onset of the nonlinear FRA regime. Meanwhile, as shown in the Fig.~\ref{LG}(d), the pump depletion primarily occurs in the outer region, corresponding to the energy transfer to the LG seed beam that has doughnut-liked transverse electric field profile. In the final stage ($\sim 190T_1$), the amplified seed pulse continues to  propagate in the plasma. Since its intensity is already high enough to trigger self-compression, the seed pulse will be self-compressed to an ultra-short duration with a further enhancement in the peak intensity. As shown in Fig.~\ref{LG}(e)$\sim$(f), the seed pulse is amplified from an initial intensity of $1\times 10^{12}\,\mathrm{W/cm^2}$ to an output intensity of $1\times 10^{17}\,\mathrm{W/cm^2}$, with its duration compressed from $90\,\mathrm{fs}$ to $10\,\mathrm{fs}$. The helical structure of the output seed pulse confirms retention of its initial OAM mode ($l_1=1$).

To further demonstrate the scalability of the FRA scheme on LG beams amplification,  we then vary the OAM mode of the seed pulse and the polarization state of the pump pulse. Figure \ref{LG_l}(a) and \ref{LG_l}(b) illustrate the transverse electric field distribution $E_y$ of the output seed pulse with initial OAM mode $l_{1y}=2$ and $l_{1y}=3$, respectively. These 3D PIC simulation results confirm that the FRA can effectively amplifies LG beams with different OAM modes, achieving output intensities around $10^{17}\,\mathrm{W/cm^2}$ while preserving the initial OAM mode. We also explore the generation and amplification of new OAM modes in the seed pulse. In this case, a circularly polarized super-Gaussian pump pulse without OAM is adopted to pump a $y-$linearly polarized seed LG pulse carrying an initial OAM mode of  $l_{1y}=1$. The simulation reveals that while the original OAM mode in the $E_y$ component is amplified (Fig.~\ref{LG_l}(c)), a new OAM mode emerges and is amplified in the orthogonal $E_z$ component for the seed LG pulse (Fig.~\ref{LG_l}(d)). The $E_z$ field distribution shows a single-helical structure, confirming the acquisition of new OAM mode of $l_{1z}=1$. This result is consistent with the theoretical analysis given by Eq.~\eqref{a1_OAM}.

\subsection{Bessel-Gaussian Beams}

In this subsection, we consider the amplification of a Bessel-Gaussian (BG) pulse via FRA.  The electric field of the BG pulse can be described as~\cite{durnin1987diffraction,gori1987bessel}
\begin{align}
a_{\mathrm{BG}}=a_0J_0(\frac{r}{r_0})\mathrm{exp}\left(-\frac{r^2}{2\sigma_0^2}+\phi_0(x,t)\right),
\end{align}
where $a_0$ is the normalized laser vector potential, $J_0(r/r_0)$ is the zeroth-order Bessel function of the first kind, with $r$ denoting the radial distance in the transverse plane and $r_0$ the full width at half maximum of the central peak. Here, $\sigma_0$ represents the transverse Gaussian envelope that ensures the finite energy of the Bessel beam, and the paraxial approximation is considered. Compared with a regular Gaussian beam, the presence of the term $J_0(r/r_0)$ in the beam profile leads to the transverse amplitude modulation that the BG pulse has the secondary symmetric peaks along transverse direction.

\begin{figure}[t]
\centering
\includegraphics[width=1.0\textwidth]{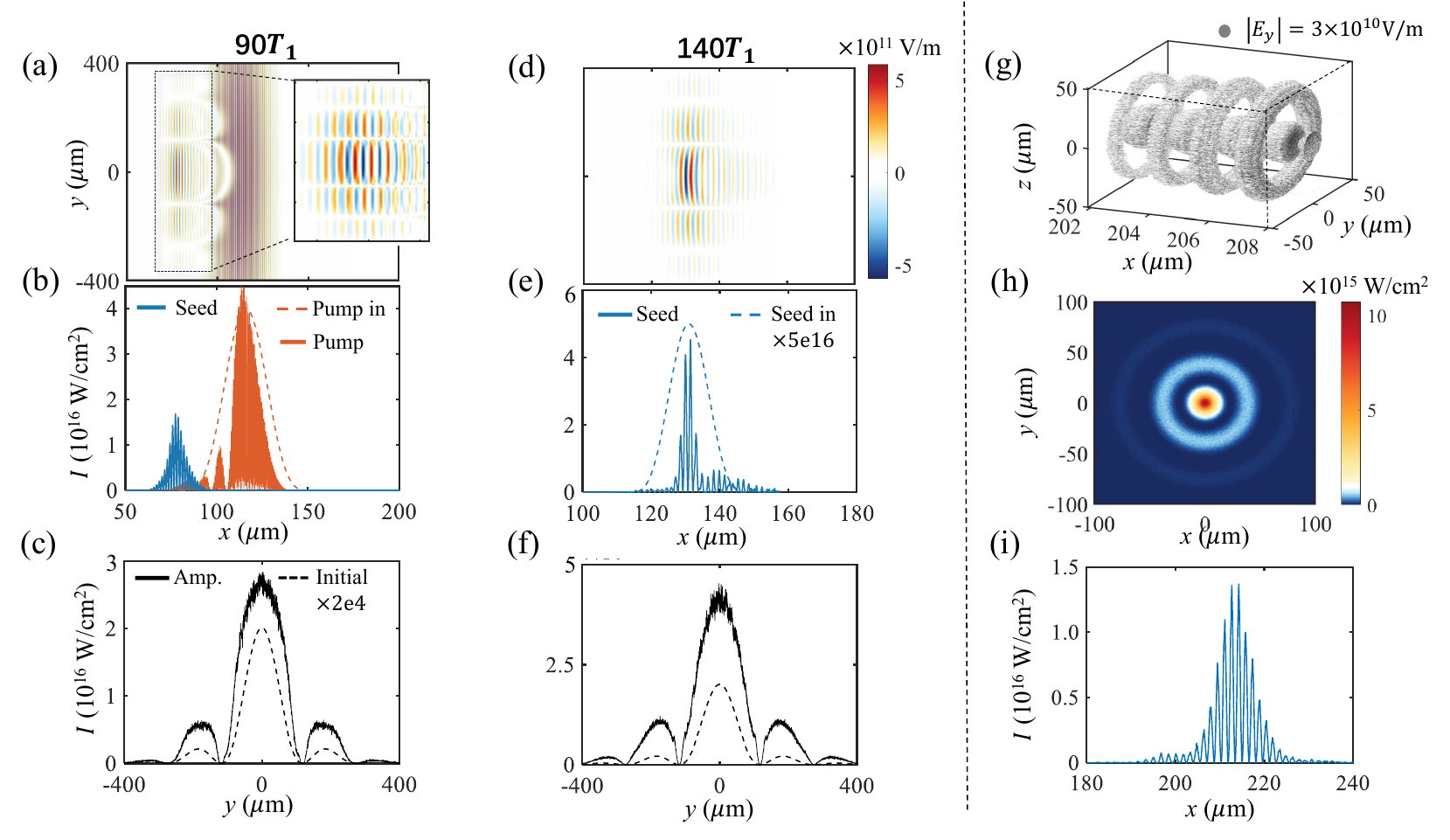}
\caption{\textbf{Simulation results of the BG beam amplification by a Gaussian pump pulse.} (a)$\sim$(f) 2D PIC results where each column represents different time stages, each row represents the 2D spatial distributions of electric fields of pulses, the longitudinal field intensity distributions, and the transverse field intensity distributions of the seed, respectively. The dashed lines in (b) and (e) are the initial envelopes of the seed pulse (blue one, magnified by $5\times 10^{16}$), and the pump pulse (red one). (g)$\sim$(i) 3D PIC results, which are the 3D isosurfaces of the amplified BG beam electric field, 2D slice transverse intensity distributions, and 1D longitudinal intensity distributions, respectively. }
\label{BG}
\end{figure}

The PIC simulation results of BG pulse amplification via the FRA scheme in two-dimensional (2D) and 3D geometries are shown in Fig.~\ref{BG}. In the 2D case (left panels), the input seed pulse is a BG beam with wavelength $\lambda_1=1.8\,\mathrm{\mu m}$, peak intensity $I_1=1\times 10^{12}\,\mathrm{W/cm^2}$, duration $\tau_1=60\,\mathrm{fs}$, and spot size $\sigma_1=400\,\mathrm{\mu m}$ (central peak spot size $r_1=50\,\mathrm{\mu m}$). The pump pulse is a typical Gaussian beam with wavelength $\lambda_0=1.0\,\mathrm{\mu m}$, peak intensity $I_0=4\times 10^{16}\,\mathrm{W/cm^2}$, duration $\tau_0=120\,\mathrm{fs}$, and spot size $\sigma_0=800\,\mathrm{\mu m}$. Both the pump and the seed pulses have a $\sin^2$ temporal profile, and co-propagate along the $x-$direction. The plasma is uniformly distributed along $x$, with a density of $2.2\times 10^{20}\,\mathrm{cm^{-3}}$ to satisfy the three-wave coupling condition for forward Raman scattering. 

As shown in Fig.~\ref{BG}(a)$\sim$(c), at the middle stage where the seed pulse just slides out of the pump pulse, the BG seed beam is amplified to $1\times 10^{16}\,\mathrm{W/cm^2}$, and maintains its transverse profile with secondary symmetric peaks. The rear part of the pump pulse exhibits significant depletion with a pattern of wave trains (also called as "$\pi$" pulse), indicating the onset of the nonlinear amplification regime. In the final stage, as the amplified BG beam propagates further in the plasma, its intensity is high enough to trigger the self-phase modulation, compressing the pulse duration to $10\,\mathrm{fs}$ (3 cycles) and further enhancing the peak intensity to $5\times 10^{16}\,\mathrm{W/cm^2}$, as illustrated in Fig.~\ref{BG}(d)$\sim$(f).

We also carry out a 3D PIC simulation to verify the FRA efficacy for BG pulses. The basic simulation parameters are identical to these in the 2D cases, except that due to the limited size of 3D simulation box, the seed BG beam has a reduced spot size $\sigma_1=80\,\mathrm{\mu m}$ (central peak spot size $r_1=10\,\mathrm{\mu m}$), and the pump pulse has $\sigma_0=90\,\mathrm{\mu m}$ with an initial peak intensity of $2\times 10^{16}\,\mathrm{W/cm^2}$. The results are shown in the right panels of Fig.~\ref{BG}. It is found that the output seed BG pulse is amplified $10^4$ times in its intensity, and the duration is compressed while maintaining a  well-defined profile in both transverse and longitudinal dimensions.

\subsection{Airy Beams}

In the final, we apply the FRA scheme to amplify the Airy laser beams. In two-dimensional geometry, the electric field of an Airy beam propagating along the $x$ direction is given by~\cite{siviloglou2007accelerating}
\begin{align}
a_{\mathrm{Airy}}=a_0\mathrm{Ai}\left(\frac{y}{y_0}\right)\mathrm{exp}\left(\frac{\beta y}{y_0}+\phi_0(x,t)\right).
\end{align}
Here, $a_0$ is the normalized laser vector potential, $\mathrm{Ai}(y/y_0)$ is the Airy function with $y_0$ representing the transverse scale, and the parameters $\beta$ is a small positive decay constant to ensure that the Airy wave has finite energy. The transverse intensity profile of an Airy beam features a primary peak followed by a series of smaller secondary peaks exhibiting exponential decay. The Airy beam owns some unique features that its main intensity peak undergoes continuous transverse displacement and forms a curved trajectory during its propagation in free space. This process is alike to the gravity acceleration process characterized by $g=1/2k^2y_0^3$~\cite{siviloglou2008ballistic}. It is noteworthy that the above feature causes the direction of the laser wave vector to change continuously during propagation. As a result, the effective amplification of an Airy beam can only be sustained within limited spatial and temporal scales, which exactly manifests the inherent advantage of the FRA scheme.

\begin{figure}[b]
\centering
\includegraphics[width=1.0\textwidth]{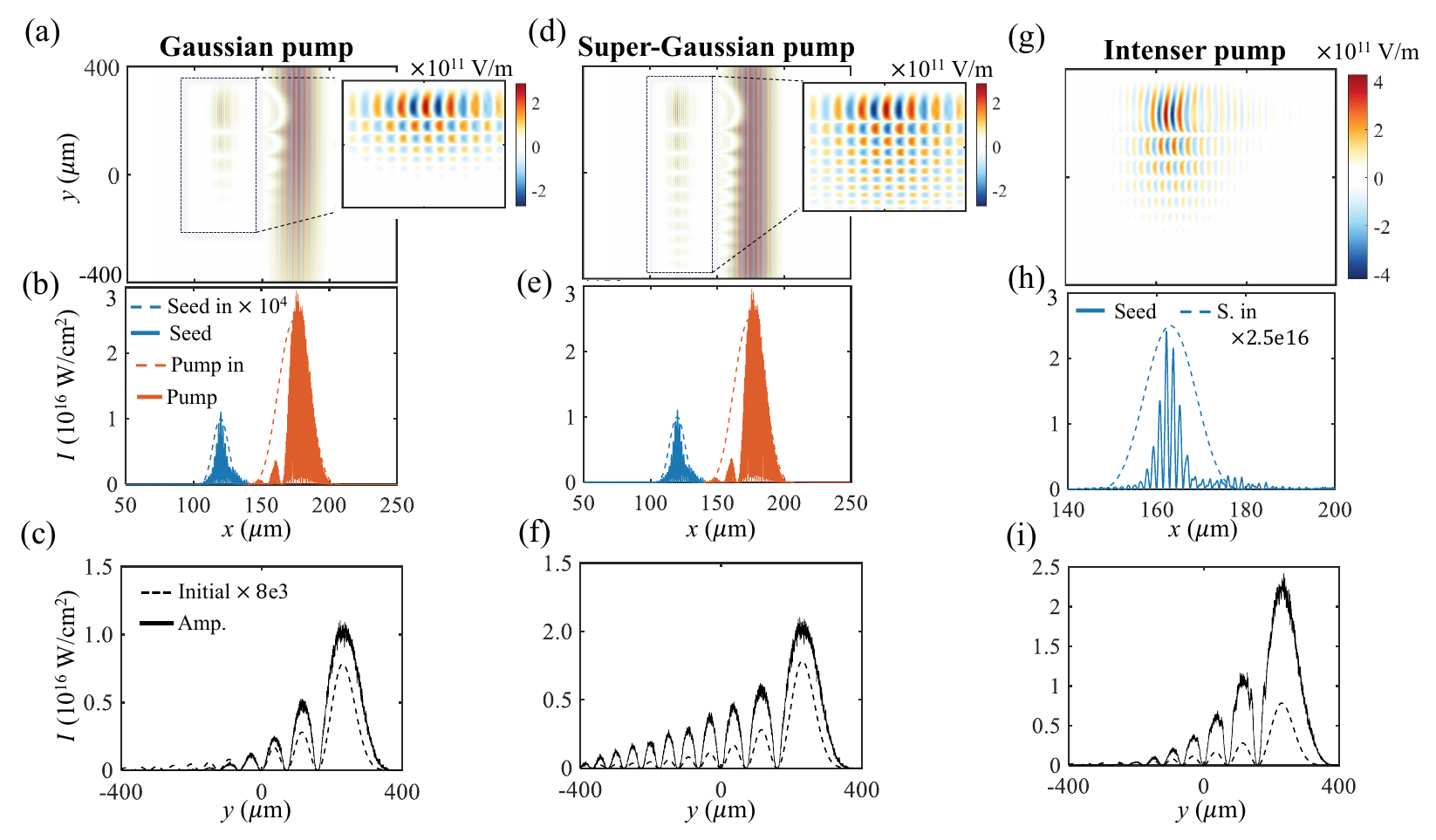}
\caption{\textbf{2D PIC Simulation results showing the amplification of an Airy seed beam by a Gaussian or super-Gaussian pump pulse.} Each row represents the 2D spatial distributions of electric fields of pulses, the longitudinal field intensity distributions, and the transverse field intensity distributions of seed, respectively. (a)$\sim$(c) Amplification by a Gaussian pump pulse. (d)$\sim$(f) Amplification by a super-Gaussian pump pulse. (g)$\sim$(i) Amplification by much intenser Gaussian pump pulse.}
\label{Airy}
\end{figure}

The 2D PIC simulation results on Airy beam amplification are presented in Fig.~\ref{Airy}. Here, the parameters of the input seed Airy pulse including initial peak intensity, wavelength, and duration are identical to those of the BG beams. Other parameters are given as $y_1=50\,\mathrm{\mu m}$, $\beta=0.1$. The pump pulse has a wavelength $\lambda_0=1.0\,\mathrm{\mu m}$, peak intensity $I_0=2.5\times 10^{16}\,\mathrm{W/cm^2}$, duration $\tau_0=120\,\mathrm{fs}$, and spot size $\sigma_0=800\,\mathrm{\mu m}$ for Gaussian transverse profile and $\sigma_0=500\,\mathrm{\mu m}$ for super-Gaussian transverse profile, respectively. The background plasma density is the same as that adopted in previous sections. As shown in the first two columns of Fig.~\ref{Airy}, in both Gaussian pump and super-Gaussian pump cases, the seed Airy beam is amplified to $10^{16}\,\mathrm{W/cm^2}$, and retains a series of decreasing peaks in the transverse field distribution. Meanwhile, the rear part of the pump pulse is significantly depleted and modulated into a wave train, corresponding to the nonlinear stage of FRA. Further, it is obvious that the super-Gaussian pump can achieve more sufficient amplification on Airy beams, and result in a greater number of well-defined secondary peaks as illustrated in the Fig.~\ref{Airy}(c) and (f). In the above two cases, the temporal compression of the seed pulse duration is not obvious. This can be attributed to two factors. Firstly, the initial seed pulse duration is already as small as the reciprocal of growth rate $g_{\mathrm{FRA}}^{-1}$, thereby the self-compression effect during the FRA is weakened. Secondly, the output seed pulse is not intense enough to trigger the efficient self-phase modulation.  When increasing the pump initial peak intensity to $3\times10^{16}\,\mathrm{W/cm^2}$, the obvious self-compression of the amplified  Airy beam is triggered. As a result, the amplified Airy beam is finally compressed to a sub-cycle duration with its peak intensity rising to $2.5\times 10^{16}\,\mathrm{W/cm^2}$ as shown in the third column of Fig.~\ref{Airy}.

\section{Discussion}
\begin{figure}[b]
\centering
\includegraphics[width=0.55\textwidth]{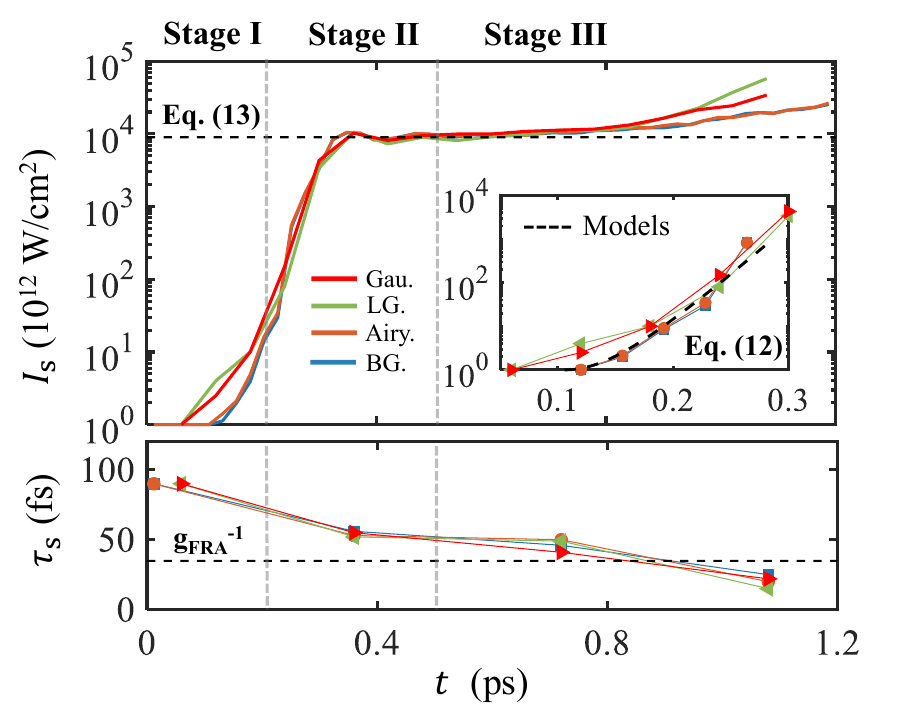}
\caption{\textbf{The amplified seed pulses peak intensity (upper one) and duration (lower one) as a function of time}, including Gaussian beam (red), LG beam (green), Airy beam (brown), and BG beam (blue). The insert plot is the a closeup of the seed peak intensity evolution at the early stage. All the black dashed lines are given by the theoretical models mentioned above.}
\label{track}
\end{figure}

To further demonstrate the effectiveness and universality of the FRA scheme applied to various spatially-structured laser beams, Fig.~\ref{track} tracks the time evolutions of the seed peak intensity (upper panel) and duration (lower panel) during the FRA of four different types of spatially-structured seed pulses, including the LG, BG, Airy and Gaussian beams. Among them, the Gaussian and LG beam results are obtained from 3D PIC simulations, while the BG and Airy beam results are from 2D PIC simulations. In each case, the laser wavelengths and the plasma density are identical to those in the above sections. All pump pulses maintain the identical parameters: duration of $180\,\mathrm{fs}$, peak intensity of $2\times 10^{16}\,\mathrm{W/cm^2}$, and spot sizes sufficiently large to fully encompass the seed pulses. All seed pulses have the same duration $90\,\mathrm{fs}$, the same initial peak intensity $1\times 10^{12}\,\mathrm{W/cm^2}$, and the spot-size as defined in the above subsections.

As shown in the Fig.~\ref{track}, it is found that the amplified seed beams exhibit similar temporal evolution patterns even if their spatial structure modes are distinctly different, and the FRA processes can be divided into three stages. In the linear stage (Stage \uppercase\expandafter{\romannumeral1}), the seed pulse grows exponentially during amplification, which is well predicted by the analytical model given by Eq.~\eqref{l_model}, as illustrated in the insert plot of the upper panel. As the pre-amplified seed pulse continues  to propagate together with the pump pulse in the plasma, its evolution enters the nonlinear stage (Stage \uppercase\expandafter{\romannumeral2}), where its intensity grows even more rapidly. At the end of this stage, the seed intensity saturates around $1.5\times 10^{16}\,\mathrm{W/cm^2}$, which is consistent with the quantitative scaling relation given by Eq.~\eqref{nl_model}. Notably, slight fluctuation around the saturation value is observed in the amplified seed intensity, which can be attributed to the energy reverse conversion that has been well elucidated by our nonlinear model. Furthermore, the seed duration after the nonlinear stage decreases to approximately $g_{\mathrm{FRA}}^{-1}$, which is the result of the dynamics amplification of the FRA described in Sec.~\ref{sec_m}. Such pulse ”compression” effect is remarkable in the SRS amplification schemes with higher growth rates. In the final stage, the duration of the amplified seed is self-compressed to sub-cycle level via the self-phase modulation, thereby achieving a further enhancement of the pulse peak intensity.

We also compare our FRA scheme with the two other well-known plasma-based light amplification schemes, i.e. backward Raman scattering~\cite{vieira2016amplification} and strong-coupling backward Brillouin scattering~\cite{wu2024efficient}, on the amplification of LG laser pulses. As shown in Table.~\ref{tab1}, the FRA scheme can amplify LG pulses with longer wavelengths while achieving comparable amplification efficiency with shorter pump pulse durations, even when the initial seed pulse is much weaker. Furthermore, the self-phase modulation effect can compress the seed pulse within plasma density regimes applicable to the FRA scheme. Overall, our proposed FRA scheme is designed to efficiently amplify initially extreme weak laser pulses with much longer wavelengths to high intensities. The entire amplification process can be achieved within short temporal scales (a few hundred femtoseconds) and compact plasma lengths (a few hundred micrometers). The high-density plasma and the co-propagating configuration enable more sufficient interaction between the pump and seed even with limited pulse duration while maintaining the high energy transfer efficiency. This enables one to employ a pump with a relatively short duration for amplification, which brings advantages such as relieving the plasma heating effects, reducing the length of the uniformly distributed plasma, and depressing the instability growth. These features should make the FRA scheme more accessible in the experiments and more applicable in practical implementations.

\begin{table}[h]
\caption{Comparison of different light amplification schemes on LG beams, reported in Refs.~\cite{wu2024efficient}, ~\cite{vieira2016amplification}.}\label{tab1}%
\begin{tabular}{@{}lllllllll@{}}
\toprule
Schemes\footnotemark[1]  & $\lambda_s/\lambda_p$ & $I_{p0}$ & $T_{p0}$ & $I_{s0}$ & $T_{s0}$ & $I_{s,\mathrm{out}}$ & $T_{s,\mathrm{out}}$ & $r$
\\
& & $(\mathrm{W/cm^2})$ & $(\mathrm{fs})$ & $(\mathrm{W/cm^2})$ & $(\mathrm{fs})$ & $(\mathrm{W/cm^2})$ & $(\mathrm{fs})$ & $(\mathrm{\mu m})$
\\
\midrule
B-SBS~\cite{wu2024efficient} & $1.0$   & $10^{16}$         & $800$  & $5\times 10^{15}$ & $100$ & $10^{17}$         & $60$ & $12$ \\
B-SRS~\cite{vieira2016amplification} & $1.053$ & $10^{15}$         & $2500$ & $7\times 10^{15}$ & $25$  & $3\times 10^{17}$ & - & $400$ \\
F-SRS (this work) & $1.8$   & $2\times 10^{16}$ & $180$  & $10^{12}$         & $90$  & $10^{17}$         & $13$ & $40$ \\
\botrule
\end{tabular}
\footnotetext[1]{All three are 3D PIC simulation results. B-/F- before SRS and SBS refers to the backward/forward scattering. The index p/s refers to the pump pulse and seed pulse, and the index 0/out refers to the pulse initial state and output state after amplification.}
\end{table}

\section{Conclusion}
We have proposed a new scheme to efficiently amplify spatially-structured laser beams via forward Raman amplification, in which a Gaussian pump laser pulse and a spatially-structured seed pulse co-propagate in plasma. An analytical model based on the three-wave coupling equations is developed, showing that: (1) Under the paraxial approximation, the FRA scheme can effectively amplify the seed pulse with any arbitrary transverse profile; (2) The OAM conservation is maintained in the FRA; (3) The presented linear and nonlinear models are universally applicable to the amplification of various spatially-structured laser beams, as verified by the PIC simulations. Through 2D and 3D PIC simulations, the FRA scheme is demonstrated to be applicable for LG, BG, and Airy beams, where an initial seed pulse at around $10^{12}\,\mathrm{W/cm^2}$ can be amplified $10^4\sim 10^5$ times to the intensities exceeding $10^{16}\,\mathrm{W/cm^2}$ within a short distance in plasma. Moreover, the amplified seed pulse can be self-compressed to nearly a single optical cycle, with its intensity further boosted while suppressing the instabilities. This provides a feasible way towards the generation of extremely high-power spatially-structured lasers: first, generating a weak seed pulse via conventional optics elements, then amplifying it via the FRA in plasma. It is worth mentioning that the amplification of BG and Airy beams studied in this work have not been explored in any other plasma-based amplification schemes. These two spatially-structured laser beams exhibit some unique propagation properties in free space. Their amplification requires an ultrashort interaction distance, which may be achievable solely via the FRA scheme. The FRA scheme is expected to be applicable for other spatially-structured laser beams such as Hermite–Gaussian (HG) or vector beams.

Moreover, our scheme operates effectively across a wide range of laser and plasma parameters, demonstrating unique advantages for amplifying spatially-structured infrared light to extremely high power. Compared with existing light amplification schemes such as CPA and OPCPA in crystals or backward Raman and Brillouin amplification in plasma, our FRA scheme has the following distinct merits. (1) \textbf{PW-Class Output:} By employing a large laser spot size (e.g., mm-scale), the FRA scheme enables the generation of PW-class near-infrared spatially-structured lights. (2) \textbf{Near-Single-Cycle Durations:} The amplified seed pulses can be self-compressed to nearly single-cycle durations without significant development of instabilities, making them particularly suitable for applications such as high harmonics generation, attosecond pulse generation, and ultrafast science. (3) \textbf{High-Repetition-Rate Operation:} The use of plasma as the amplification medium enables the scheme to operate at a high repetition rate, and the seed pulse can be amplified to high intensity directly without the need of post-compression. (4) \textbf{High Temporal Contrast:} Our scheme may enable the amplified pulse to have an ultrahigh temporal contrast ratio, addressing a key challenge to conventional chirped pulse amplification in crystals. (5) \textbf{Simplified Experimental Implementation:} The configuration with co-propagating seed and pump laser pulses enables relatively easy implementation in experiments. (6) \textbf{Cascadability for Mid-Infrared Wavelengths:} By adjusting the plasma density, the cascaded amplification towards longer wavelengths in the mid-infrared regime becomes possible. For example, a $1.0\,\mathrm{\mu m}$ pump pulse can first amplify a $1.8\,\mathrm{\mu m}$ seed pulse, which can subsequently serve as a pump pulse to amplify a secondary seed pulse with longer wavelength of $3.3\,\mathrm{\mu m}$, and so on. (7) \textbf{Two-Color Light Source:} The amplified seed pulse and the remaining pump pulse may be jointed deployed as a novel two-color light source for advanced applications. In a word, the proposed FRA scheme offers new opportunities for ultrafast science and high-field physics.

\section*{Acknowledgments}
This work is supported by the National Natural Science Foundation of China (Grant Nos. 12135009, 12005287, and 12375236) and by the Strategic Priority Research Program of Chinese Academy of Sciences (Grant Nos. XDA25050100 and XDA25010100). Numerical simulations were performed on Computer $\pi$2.0 in the Center for High Performance Computing at Shanghai Jiao Tong University.

\bibliography{sn-bibliography}

\end{document}